\documentstyle[epsf,epsfig,preprint,aps]{revtex}
\draft
\preprint{SNUTP 01/009}
\begin{document}
\title{\Large\bf Effective supersymmetric theory and
muon anomalous magnetic moment with $R$ parity violation} 
\author{Jihn E. Kim,\footnote{jekim@phyp.snu.ac.kr}
Bumseok Kyae\footnote{kyae@fire.snu.ac.kr} and Hyun Min 
Lee\footnote{minlee@phya.snu.ac.kr}} 
\address{ Department of Physics and Center for Theoretical
Physics, Seoul National University,
Seoul 151-747, Korea}
\maketitle
\begin{abstract}
The effective supersymmetric theory (ESUSY) with $R$ parity 
conservation cannot give a large anomalous magnetic moment of $\mu$. 
It is pointed out that the flavor conservation and a large 
$(g-2)_\mu$ within the experimental limits are achievable in the 
ESUSY with $R$ parity violating couplings involving the third 
generation superparticles.
\end{abstract}
[Key words: anomalous magnetic moment of muon, effective
SUSY, $R$ parity violation]

\newpage

\def\smn{$\sigma_{\mu\nu}$}
\def\p{\partial}

The recent hint~\cite{bnl} that the anomalous magnetic moment of muon $a_\mu
=\frac{1}{2}(g-2)_\mu$ might not fall within the predicted range in
the standard model attracted a great deal of attention in the hope of 
probing new physics possibility~\cite{susy,techni}. Probably,
in the future this excitement can be compared to that on the discussion 
of the weak neutral current thirty years ago~\cite{nc}.
  
In general, the loop correction to the magnetic moment of $\mu$ arises
if there exists a left to right (or right to left) chirality 
transition of the external muon lines.
This chirality transition needs an insertion of a 
fermion mass or a Yukawa coupling vertex. If heavy fermions are
introduced, it has been known that a large magnetic moment is
possible, with both left-handed and right-handed currents~\cite{cl}.
Without a heavy fermion, the mass insertion or the Yukawa coupling
leads to a magnetic moment proportional to the external
fermion mass. But in the standard model there does not exist
a heavy fermion and we expect small
lepton magnetic moments of order $a_\nu\sim 
(m_\nu m_e/M_W^2)\sim 10^{-18}$ times the electron 
Bohr magneton~\cite{cl} for the eV range neutrino mass and 
$a_\mu\sim (m_\mu^2/M_W^2)$ times the
muon Bohr magneton~\cite{magsm}. There is a new heavy(but relatively light
to resolve the hierarchy problem)
fermion in supersymmetric models, i.e. the chargino.
Then the above argument may imply an anomalously large anomalous
magnetic moment of $\mu$, which however is not realized due
to the chiral nature of the supersymmetric models~\cite{susy}.

Indeed, the recent observation of the
anomalous magnetic moment of muon are around the electroweak
scale order~\cite{bnl}, i.e. of order predicted in the
standard model~\cite{magsm}
\begin{equation}
a_\mu^{\rm exp}-a_\mu^{\rm th} = 426(165)\times 10^{-11}, 
\end{equation}
but off from the standard model prediction by $2.6\sigma$. 
The theoretical prediction quoted above is from Ref.~\cite{davier}
which gives the most stringent error bar compared to
the other published results~\cite{yind}, the
main difference coming from the treatment of the
hadronic contributions. Therefore, at present it is premature
to conclude the existence of a new physics beyond 100 GeV.

Nevertheless, it is tempting to search for possibilities of
generating $(g-2)_\mu$ of order the electroweak scale or 
even a larger value. Then, the 
mass scale of new physics must be close to the
electroweak scale since the possible deviation is of order
the predicted range in the standard model.
In this spirit, already there appeared several 
explanations~\cite{susy,techni}.
Among these, the supergravity scenario is particularly interesting
because a relatively light (mass $<$ 500 GeV) superpartners are 
around the corner as shown in~\cite{susy}, 
which gives a hope of probing whole
spectrum of superparticles below TeV at TevatronII and LHC.

However, the supergravity models have the notorious problem regarding
the flavor changing processes~\cite{fcnc}. One of the models resolving
this flavor changing problem is the gauge mediated supersymmetry breaking
at low energy~\cite{fcnc}. 
Another model is the effective supersymmetric theory(ESUSY)
in which the superpartners of the first two generations are heavy
(masses greater than tens of TeV)
while the superpartners of the third generation is
relatively light (masses a few hundred GeV)~\cite{cohen}. 
Thus, the ESUSY
with $R$ parity conservation cannot account for the possible
extra contribution to the $g-2$ of muon, reported recently~\cite{bnl},
since smuonneutrino($\tilde\nu_\mu$) and smuon
($\tilde \mu$) masses are very heavy $>20$ TeV in ESUSY.

In this paper, we consider both the flavor changing problem and the
BNL experiment~\cite{bnl} seriously, and study
the anomalous magnetic moment of
$\mu$ in the ESUSY. Here, we do not
digress into how the ESUSY phenomenon results at low energy,
but mainly discuss its effects on the $(g-2)_\mu$. In the effective
SUSY scenario, sparticle masses of the first two generations 
are required to be heavier than 20 TeV~\cite{cohen}.
These sparticles of the first two generations are not
responsible for a large anomalous magnetic moment of
$\mu$. Therefore, the SUSY loops considered in
Ref.~\cite{susy} for $a_\mu$ are negligible in ESUSY. For 
the gauge hierarchy solution, however, there is an effective
SUSY for the third generation particles. The phenomenological constraints
on the third generation sparticle masses are
expected to be lighter than 1 TeV~\cite{cohen}.
Within the experimentally allowed regions of these mass parameters,
we look for a possibility of {\it generating a large anomalous magnetic
moment of $\mu$ by introducing $R$ parity violating interactions.}

\def\mstop{M_{\tilde t}}
\def\mstopl{M_{\tilde t,L}}
\def\mstopr{M_{\tilde t,R}}
\def\msbottom{M_{\tilde b}}
\def\msbottoml{M_{\tilde b,L}}
\def\msbottomr{M_{\tilde b,R}}
\def\mstau{M_{\tilde \tau}}
\def\mstaul{M_{\tilde \tau,L}}
\def\mstaur{M_{\tilde \tau,R}}
\def\msnu{M_{\tilde \nu}}
\def\msnul{M_{\tilde \nu_\tau,L}}

The ESUSY requires that top squarks must be lighter than 1 TeV
for a hierarchy solution and $SU(2)\times U(1)$ breaking by the
top quark Yukawa coupling~\cite{ir}: the left-handed stop mass 
$\mstopl <$ 1 TeV and the right-handed stop mass $\mstopr < 1$ TeV. 
Because of the $SU(2)$ symmetry left-handed sbottom mass $\msbottoml$ 
must be smaller than 1 TeV, while right-handed sbottom mass 
$\msbottomr$ need not to be constrained to $<1$ TeV. Similarly,
$\msnul, \mstaul$ and $\mstaur$ are not restricted. However, 
the Fayet-Iliopoulos anomaly should lead to a light Higgs boson
below 20 TeV for the $SU(2)\times U(1)$ breaking, i.e. there
remains an effective SUSY for the third generation below 20 TeV, 
which gives a condition~\cite{cohen}
\begin{equation}
{\rm Tr}~YM^2_{\rm heavy}=0,\label{trace}
\end{equation}
where $Y$ is the electroweak hypercharge.
If the full third generation sparticles are light, the above trace is
can be satisfied under the assumption that the first
two generations satisfy the trace rule. In addition, note that if the
third generation scalar leptons are heavy with the same mass
and the third generation scalar quarks are light then Rule
(\ref{trace}) can also be satisfied, which however is not considered
below. 
Therefore, for simplicity and in search of a possibility, we
{\it assume that the full third generation masses are light with mass
less than 1 TeV.} 

In the estimation of the anomalous magnetic
moment, there can be a mixing between the left scalar and
the right scalar fermions, which however gives a contribution
negligible compared to the case without the mixing.
The main reason for this is that the standard model
is a chiral theory and scalars introduced in the 
supersymmetrization remembers this chiral nature. So if there is
a mixing term between left- and right-sfermion masses, then there
must be a chirality flip by a mass insertion in the fermion line
in the loop. However, the $SU(3)\times SU(2)\times U(1)$ symmetry
forbids the mass insertion for charged leptons and quarks
as an internal fermion line. The only allowed diagram for the
internal neutrino line, which is possible only by introducing 
neutrino masses beyond the standard model, 
is as shown in Fig. 1 (b). We can anticipate
Fig. 1 (a), but it is absent due to the illegal $M_{\tilde \nu}$
insertion in the supersymmetric limit. (Note however that there will
be a small correction if we consider the $A$ term arising from 
$\lambda^\prime_{333}$ coupling which must be very small in view of
a small $\nu_\tau$ mass.) 

\vskip 0.3cm
\begin{figure}[tbh]
\centering
\centerline{\epsfig{file=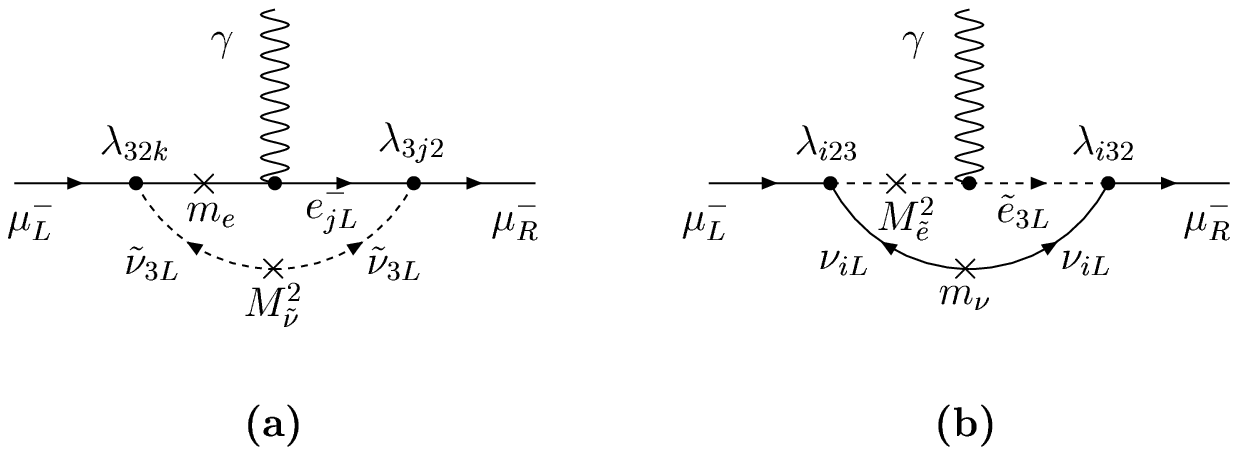,width=16cm}}
\end{figure}
\centerline{ Fig.~1.\ \it
The contribution from the slepton mixing term.}
\vskip 0.3cm

\noindent Since neutrino 
masses are below 10 eV, we can neglect Fig. 1 (b) also and hence 
in the remainder of this paper we will neglect the mass mixing 
between the left- and right-handed sfermions.
Also, for the mixing between squarks, there does not appear an
important contribution.

The relevant $R$ parity violating superpotential($\lambda$ and 
$\lambda^\prime$ couplings) is given by 
\begin{eqnarray}
W=\frac{1}{2}\lambda_{ijk}L_i L_j E^c_k
+\lambda'_{ijk}L_i Q_j D^c_k,
\end{eqnarray}
where $i,j,$ and $k$ are generation indices, 
$L_i$ and $Q_i$ are the left-handed lepton and quark 
doublet superfields, and $E^c_i$ 
and $D^c_i$ are the left-handed anti-lepton and anti-quark 
singlet superfields. The bosonic symmetry implies, $\lambda_{ijk}
=-\lambda_{jik}$.

The needed vertices for the $(g-2)_\mu$ calculation
can be read from the above superpotential,
\begin{eqnarray}
-{\cal L}_Y&=&\lambda_{ijk}(\tilde{\nu}_{iL}e_{jL}e^c_{kL}
+\tilde{e}_{jL}\nu_{iL}e^c_{kL}+\tilde{e}^*_{kR}\nu_{iL}e_{jL})\nonumber\\
&+&\lambda'_{ijk}(\tilde{\nu}_{iL}d_{jL}d^c_{kL}
+\tilde{d}_{jL}\nu_{iL}d^c_{kL}+\tilde{d}^*_{kR}\nu_{iL}d_{jL}
-\tilde{e}_{iL}u_{jL}d^c_{kL}-\tilde{u}_{jL}e_{iL}d^c_{kL}
-\tilde{d}^*_{kR}e_{iL}u_{jL})\nonumber \\
&+&h.c.
\end{eqnarray}

In terms of the four component spinors, we obtain
\begin{eqnarray}
-{\cal L}_Y\ =\ 
&\lambda_{ij2}&(\tilde{\nu}_{iL}\bar{\mu}P_L e_j
+\tilde{e}_{jL}\bar{\mu}P_L\nu_i)
+\lambda^*_{i2k}(\tilde{\nu}^*_{iL}\bar{\mu}P_R e_k
+\tilde{e}_{kR}\bar{\mu}P_R\nu^c_i)\nonumber \\ 
&-&\lambda^{\prime *}_{2jk}(\tilde{u}^*_{jL}\bar{\mu}P_R d_k
+\tilde{d}_{kR}\bar{\mu}P_R u^c_j)\nonumber \\
&+&h.c.
\end{eqnarray}
where the left-handed and right-handed chirality
projecton operator are defined as $P_{L,R}=\frac{1}{2}(1\mp\gamma^5)$.

As commented above, the slepton mixing diagrams, Fig. 1, are not
important for $(g-2)_\mu$.

The $R$ violating $\lambda$ couplings 
contribute~\cite{lev} to the muon anomalous magnetic moment as
shown in Fig.~2, $(a)-(d)$. In Figs. 2, one of the external lines 
shown as left-handed muon is understood as shifted to a right-handed
one by $m_\mu$ insertion in that external line. Then, we obtain,
\begin{eqnarray}
a^{\tilde{\nu}_\tau (1)}_\mu &=&\frac{m^2_\mu}
{16\pi^2}| \lambda_{i2k}|^2\int^1_0 dx
\frac{x^2-x^3}{m^2_\mu x^2+(m^2_{e_k}-m^2_\mu)x+m^2_{\tilde{\nu}_{iL}}
(1-x)} \nonumber\\
&\simeq&\frac{|\lambda_{32k}|^2}{48\pi^2}
\frac{m^2_\mu}{m^2_{\tilde{\nu}_{\tau L}}},\\
a^{\tilde{\tau}_R}_\mu &=&\frac{m^2_\mu}{16\pi^2}|
\lambda_{i2k}|^2\int^1_0 dx
\frac{x^3-x^2}{m^2 x^2+(m^2_{\tilde{e}_{kR}}-m^2_\mu)x+m^2_{\nu_i}(1-x)}
\nonumber\\
&\simeq&-\frac{|\lambda^*_{i23}|^2}{96\pi^2}
\frac{m^2_\mu}{m^2_{\tilde{\tau R}}},\\
a^{\tilde{\nu}_\tau (2)}_\mu &=&\frac{m^2_\mu}
{16\pi^2}|\lambda_{ij2}|^2 \int^1_0 dx
\frac{x^2-x^3}{m^2_\mu x^2+(m^2_{e_j}-m^2_\mu)x+m^2_{\tilde{\nu}_{iL}}(1-x)}
\nonumber\\
&\simeq&\frac{|\lambda_{3j2}|^2}{48\pi^2}
\frac{m^2_\mu}{m^2_{\tilde{\nu}_{\tau L}}},\\ 
a^{\tilde{\tau}_L}_\mu &=&\frac{m^2_\mu}{16\pi^2}|\lambda_{ij2}|^2\int^1_0 dx 
\frac{x^3-x^2}{m^2_\mu x^2+(m^2_{\tilde{e}_{jL}}-m^2_\mu)x+m^2_{\nu_i}(1-x)}
\nonumber\\
&\simeq&-\frac{|\lambda_{i32}|^2}{96\pi^2}
\frac{m^2_\mu}{m^2_{\tilde{\tau L}}}. 
\end{eqnarray} 
Therefore, the $\lambda$ contribution to the muon
anomalous magnetic moment is
\begin{equation}
a^{\lambda}_\mu
=\frac{m^2_\mu}{96\pi^2}\left(
|\lambda_{32k}|^2
\frac{2}{m^2_{\tilde{\nu}_\tau}}
+|\lambda_{3k2}|^2\left[
\frac{2}{m^2_{\tilde{\nu}_\tau}}
-\frac{1}{m^2_{\tilde{\tau}_L}}\right]
-\frac{|\lambda_{k23}|^2}{m^2_{\tilde{\tau}_R}}
\right)\label{l}
\end{equation}
where the sum over the repeated indices is implied.

\vskip 0.3cm
\begin{figure}[tbh]
\centering
\centerline{\epsfig{file=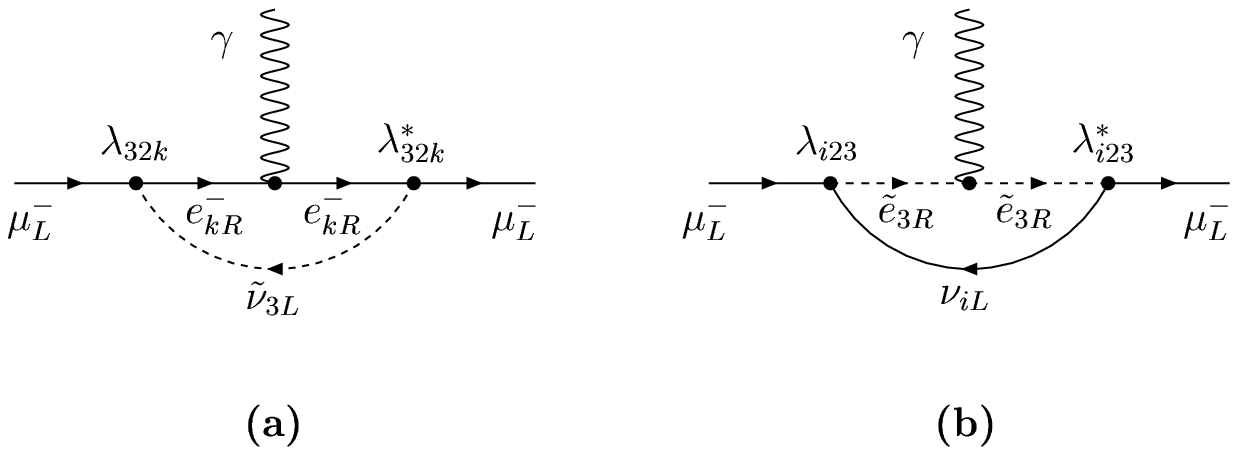,width=16cm}}
\vskip 1cm
\centerline{\epsfig{file=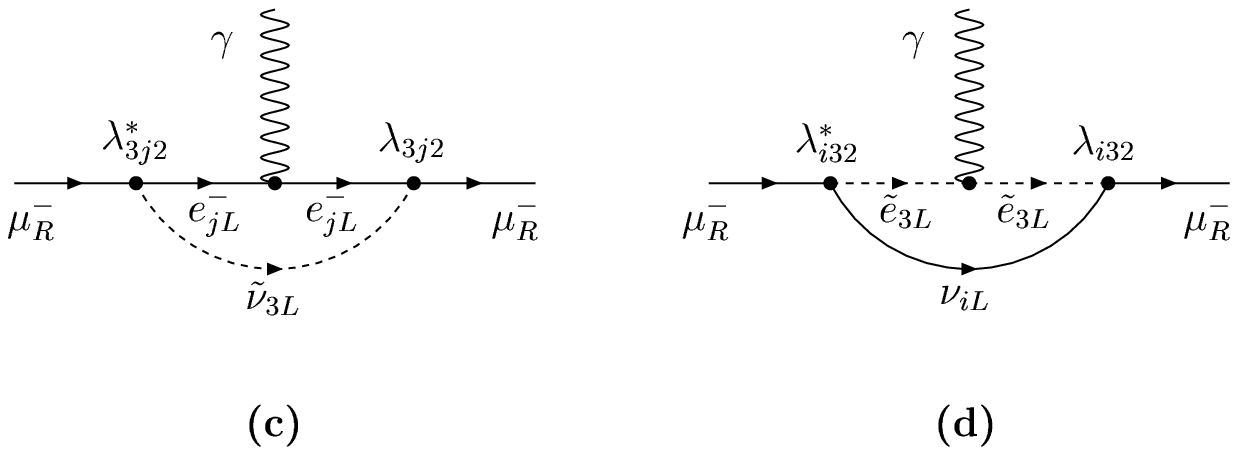,width=16cm}}
\end{figure}
\centerline{ Fig.~2.\ \it
The contribution from the $\lambda$ couplings. One $\mu_L$ (or} 
\centerline{\it 
$\mu_R$) is changed to $\mu_R$ (or $\mu_L$) 
by $m_\mu$ insertion.} 
\vskip 0.3cm

The $\lambda^\prime$ couplings also contribute to the
muon anomalous magnetic moment. The effect of soft
supersymmetry breaking term can be considered,
but we find that they are not important for our purpose.
In addition, the constraint from the 
flavor changing effects restrict possible
models, and here we concentrate on one allowed model
as discussed in the introduction, the ESUSY model; 
thus the allowed couplings
are rather restricted. Then, we obtain~\cite{lev}
\begin{eqnarray}
a^{\tilde{t}_L(1)}_\mu &=&3\bigg(\frac{1}{3}\bigg)\frac{m^2_\mu}{16\pi^2}
|\lambda^{\prime}_{2jk}|^2\int^1_0 dx
\frac{x^2-x^3}{m^2_\mu x^2+(m^2_{d_k}-m^2_\mu)x+m^2_{\tilde{u}_{jL}}(1-x)}
\nonumber\\
&\simeq&\frac{|\lambda^{\prime}_{23k}|^2}{48\pi^2}
\frac{m^2_\mu}{m^2_{\tilde{t}_L}} \\
a^{\tilde{b}_R(1)}_\mu &=&3\bigg(\frac{1}{3}\bigg)\frac{m^2_\mu}{16\pi^2}
|\lambda^{\prime}_{2jk}|^2\int^1_0 dx
\frac{x^3-x^2}{m^2_\mu x^2+(m^2_{\tilde{d}_{kR}}-m^2_\mu)x+m^2_{u_j}(1-x)}
\nonumber\\
&\simeq&-\frac{|\lambda^{\prime}_{2j3}|^2}{96\pi^2}
\frac{m^2_\mu}{m^2_{\tilde{b}_{R}}-m^2_{u_j}}(1+F_1(x_{u_j \tilde{b}_R}))
\label{mt1}\\
a^{\tilde{b}_R(2)}_\mu &=&3\bigg(\frac{2}{3}\bigg)\frac{m^2_\mu}{16\pi^2}
|\lambda^{\prime}_{2jk}|^2\int^1_0 dx
\frac{x^2-x^3}{m^2_\mu x^2+(m^2_{u_j}-m^2_\mu)x+m^2_{\tilde{d}_{kR}}(1-x)}
\nonumber\\
&\simeq&\frac{|\lambda^{\prime}_{2j3}|^2}{24\pi^2}
\frac{m^2_\mu}{m^2_{\tilde{b}_R}-m^2_{u_j}}(1+F_2(x_{u_j \tilde{b}_R}))
\label{mt2}\\
a^{\tilde{t}_L(2)}_\mu &=&3\bigg(\frac{2}{3}\bigg)\frac{m^2_\mu}{16\pi^2}
|\lambda^{\prime}_{2jk}|^2\int^1_0 dx 
\frac{x^3-x^2}{m^2_\mu x^2+(m^2_{\tilde{u}_{jL}}-m^2_\mu)x+m^2_{d_k}(1-x)}
\nonumber\\
&\simeq&-\frac{|\lambda^{\prime}_{23k}|^2}{48\pi^2}
\frac{m^2_\mu}{m^2_{\tilde{t}_L}}
\end{eqnarray}
where
\begin{eqnarray}
F_1(x)&=&\frac{6x}{1-x}\bigg[\frac{1}{2}-\frac{1}{1-x}
-\frac{x}{(1-x)^2} \ln x\bigg],\\
F_2(x)&=&\frac{3x}{1-x}\bigg[\frac{1}{2}+\frac{1}{1-x}
+\frac{1}{(1-x)^2} \ln x\bigg],
\end{eqnarray}
and $x_{u_j \tilde{b}_R}=m^2_{u_j}/m^2_{\tilde{b}_R}$.
These $F$ functions are important only for the top quark, i.e.
for $x_t\equiv x_{u_3 \tilde b_R}$. Note that $a_\mu^{\tilde t_L(1)}$
and $a_\mu^{\tilde t_L(2)}$ add up to zero. Thus, the remaining 
Eqs. (\ref{mt1}) and (\ref{mt2}) give the $\lambda^\prime$
contribution,
$$
a_\mu^{\lambda^\prime}=
\frac{m_\mu^2}{32\pi^2}\left\{
\frac{1}{m^2_{\tilde b_R}-m^2_t}
{|\lambda^\prime_{233}|^2}\left(1+\frac{2x_t}{1-x_t}\left[
\frac{1}{2}+\frac{3}{1-x_t}+\frac{2+x_t}{(1-x_t)^2}\ln x_t
\right]\right)\right\}\nonumber
$$
\begin{equation}
+\frac{m_\mu^2}{32\pi^2}\left\{
\frac{1}{m^2_{\tilde b_R}}({|\lambda^\prime_{213}|^2}
+{|\lambda^\prime_{223}|^2})\right\}>0.\label{lp}
\end{equation} 
Note that this contribution is positive definite.

\vskip 0.3cm
\begin{figure}[tbh]
\centering
\centerline{\epsfig{file=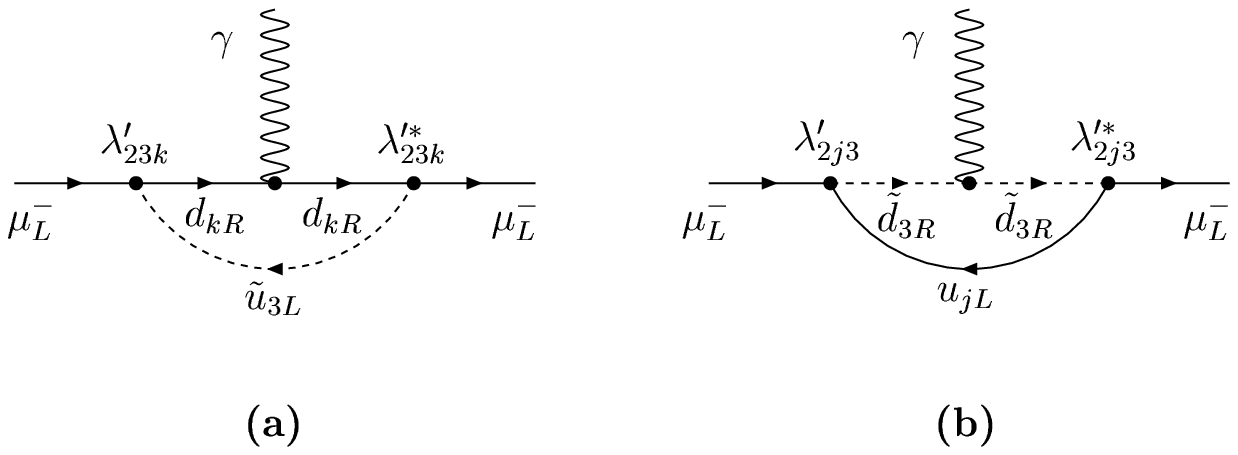,width=16cm}}
\vskip 1cm
\centerline{\epsfig{file=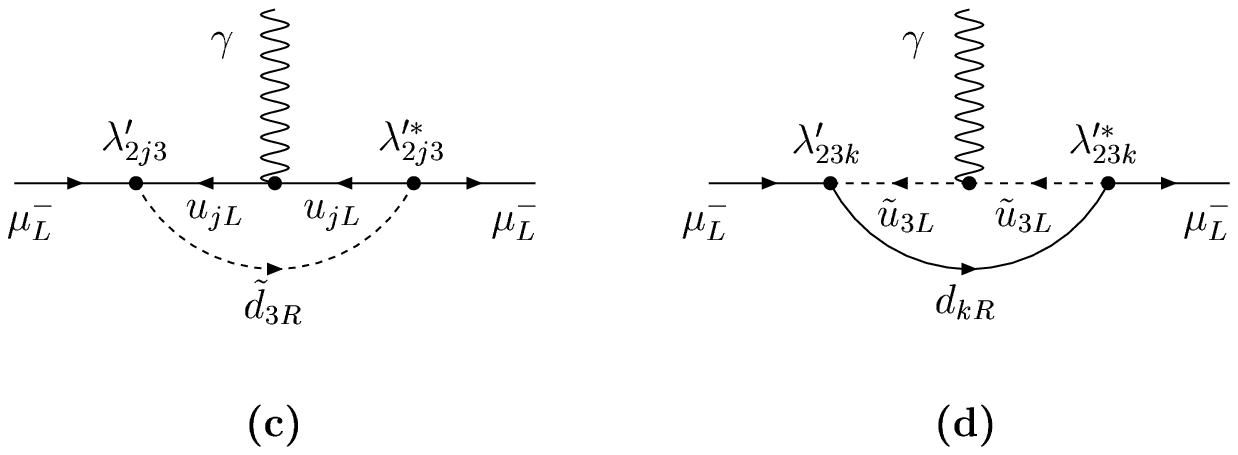,width=16cm}}
\end{figure}
\centerline{ Fig.~3.\ \it
The contribution from $\lambda^\prime$ couplings. One $\mu_L$} 
\centerline{\it 
is changed to $\mu_R$ by $m_\mu$ insertion.} 
\vskip 0.3cm

We can show from the above expression that in the 
supersymmetric limit, i.e. $\mstop=m_t$, $\msbottom=m_b$, 
etc, the contribution to $(g-2)_\mu$ vanishes, which is 
a necessary consistency check. In the expression (\ref{lp}),
the unknown mass is $m_{\tilde b_R}$. Without the
left- and right-sbottom mixing and the flavor mixing, 
it is the physical mass. 

For completeness, however, we present the formula with
the left- and right-sbottom mixing. We need 
to consider a $2\times 2$ sbottom mass matrix, in the
$(\tilde b_L,\tilde b^*_R)^T$ basis,
\begin{equation}
M^2=\left(\matrix{M_{11}^2,\ M_{12}^2\cr
M_{21}^2,\ M_{22}^2}\right).
\end{equation} 
After diagonalization of $M^2$,  $(M^2)^{-1}_{22}$ should replace
$1/m^2_{\tilde b_R}$ 
\begin{equation}
\left(V^\dagger (M_D^2)^{-1}V\right)_{22}=
\sum_k\frac{|V_{k2}|^2}{M_k^2}=\frac{1}{m^2_{\tilde b_R}}
\left[1-\frac{m_b^2(A_b+\mu\tan\beta)^2}{m^2_{\tilde b_R}
m^2_{\tilde b_L}}\right]^{-1}
\end{equation}
where $V$ is the diagonalizing unitary matrix, $M^2_D$ is the
diagonalized mass matrix, and the last relation is given for the
MSSM case.

Therefore, we obtain the total contribution to $\Delta a_\mu$ 
in the ESUSY with $R$ parity violating couplings as
\begin{equation}
\Delta a^{R\hspace{-1.75mm}/}_\mu
= a_\mu^\lambda+a_\mu^{\lambda^\prime} 
\end{equation}
where $a^\lambda_\mu$ and $a_\mu^{\lambda^\prime}$
are given in Eqs. (\ref{l}) and (\ref{lp}), respectively.
In principle, the anomalous magnetic moment can take
both signs. But for the most range of the parameter space,
in view of Eq.~(\ref{lp}), $\Delta a_\mu$ is positive. 
The possibility of a negative sign arises when the 
contributions of Fig. 2 (b) and (d) dominate.

\vskip 0.3cm
\begin{figure}
\epsfxsize=160mm
\centerline{\epsfbox{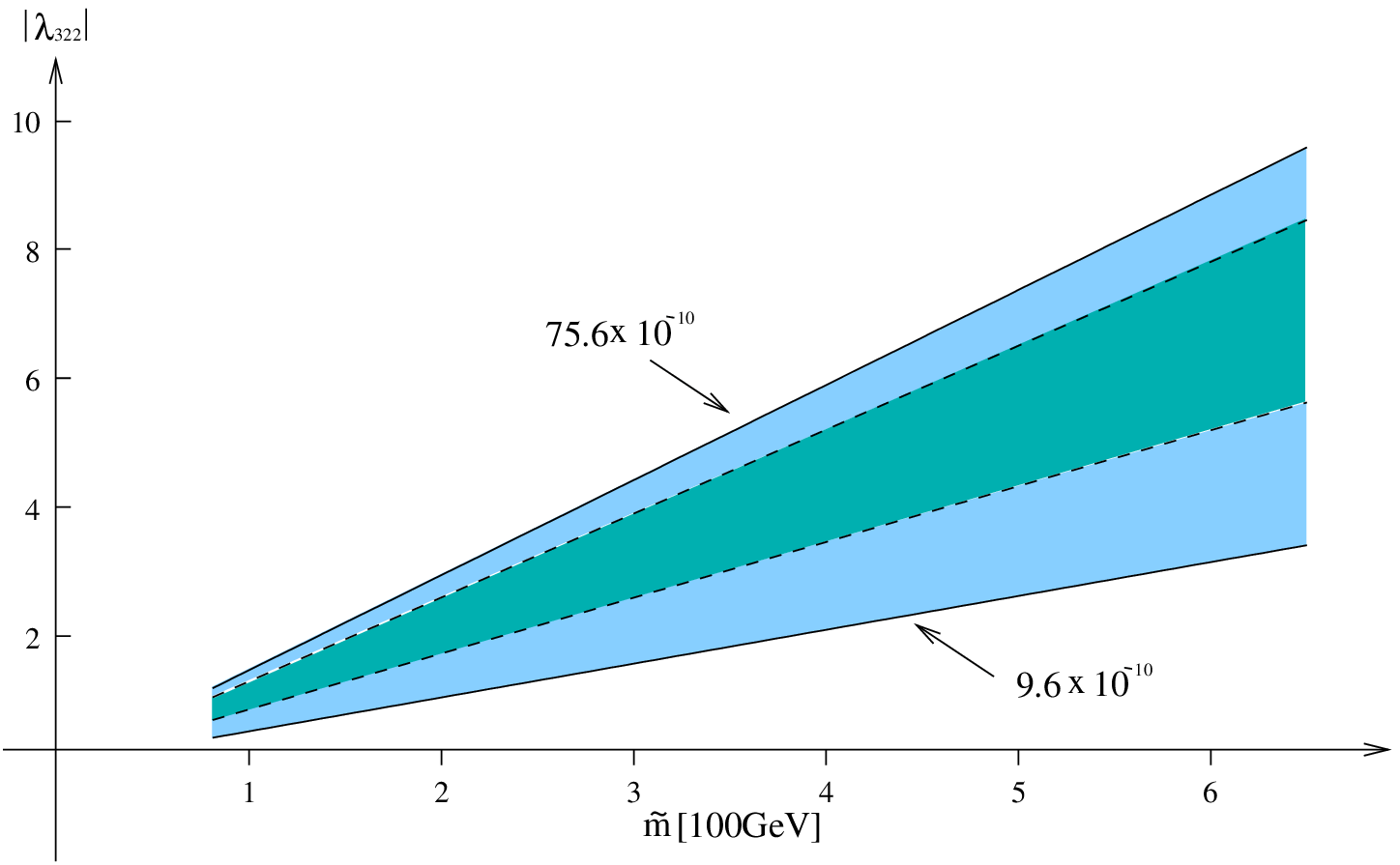}}
\end{figure}
\centerline{ Fig.~4.\ \it
The $1 \sigma$ and $2 \sigma$ band of $\Delta a_\mu$ in the
$\lambda_{322}$ and $\tilde{m}\equiv m_{\tilde{\nu}_\tau}=m_{\tilde{\tau}}$ plane.}
\centerline{\it  
The low mass cut is the LEP bound of 81.0~GeV on the stau mass~\cite{lep}.} 
\vskip 0.3cm

In view of the Brookhaven data of the anomalous magnetic moment 
of muon with the $2\sigma$ deviation\cite{bnl},
we obtain the necessary lower bound on the $R$ parity violating couplings 
($\lambda$, $\lambda'$) from Eqs.~(\ref{l}) and (\ref{lp}) as
\begin{eqnarray}
|\lambda|({\rm or}\,\,|\lambda'|)>0.524\times
\bigg(\frac{\tilde{m}}{100GeV}\bigg)
\end{eqnarray}
where $\tilde{m}$ is the third generation sfermion mass (or the sbottom mass).
In that case, the $\lambda'$ couplings cannot give sizable effects for the 
muon anomaly since there are unavoidable single bounds.  
$\lambda'_{233}$ is constrained 
from the radiative muon neutrino mass, 
$m_{\nu_\mu}\sim [3|\lambda^\prime_{233}|^2 m_b^2(A_{b}+\mu
\tan\beta)]/(8\pi^2 m^2_{\tilde b_R})$
as $|\lambda'_{233}|<0.15\times\sqrt{(m_{\tilde{b}_R}/100GeV})$
\cite{bound,chun}. 
And we also have single bounds such as 
$|\lambda'_{213}|<0.059\times(m_{\tilde{b}_R}/100GeV)$ from 
$R_\pi=\Gamma(\pi\rightarrow e\nu)/\Gamma(\pi\rightarrow \mu\nu)$ and
$|\lambda'_{223}|<0.21\times(m_{\tilde{b}_R}/100GeV)$ from the decay process 
$D\rightarrow K l\nu$\cite{bound}. 

For the $\lambda$ couplings, on the other hand, we can get a naturally large 
contribution to the muon anomalous magnetic moment in the effective SUSY model 
we are considering. Let us enumerate the $\lambda$ contribution 
from Eq.~(\ref{l}) as 
\begin{eqnarray}
a^\lambda_\mu\approx \frac{m^2_\mu}{32\pi^2 \tilde{m}^2}
\bigg(\frac{2}{3}|\lambda_{321}|^2+|\lambda_{322}|^2
+\frac{1}{3}|\lambda_{323}|^2
+\frac{1}{3}|\lambda_{312}|^2-\frac{1}{3}|\lambda_{123}|^2\bigg)
\end{eqnarray}
where we set $m_{\tilde{\nu}_\tau}=m_{\tilde{\tau}_L}
=m_{\tilde{\tau}_R}\equiv \tilde{m}$.
There also exist relevant single bounds on the $\lambda$ 
couplings. Firstly, $\lambda_{123}$ giving a negative contribution becomes 
negligible since $|\lambda_{123}|<0.05\times(m_{\tilde{\tau}_R}/100GeV)$ 
from the charged current universality\cite{bound} 
and thus the $\lambda$ contributions can be regarded 
as positive definite values consistent with the Brookhaven data 
if the other $\lambda$ couplings are allowed to be larger than 
$|\lambda_{123}|$. 
And we can also ignore $\lambda_{312}$ and $\lambda_{321}$, which just give 
contributions less than required by factor 10 since they are 
constrained as $\sqrt{|\lambda_{312}|^2+|\lambda_{321}|^2}
<0.25\times(m_{\tilde{\nu}_\tau}/100GeV)$ 
from asymmetries in $e^+ e^-$ collisions at the $Z$ peak\cite{bound}. 
Lastly, since
\begin{eqnarray}
\sqrt{|\lambda_{322}|^2
+\bigg(\frac{\tilde{M}}{\tilde{m}}\bigg)^2|\lambda_{323}|^2} 
<0.070\times\bigg(\frac{\tilde{M}}{100GeV}\bigg)\label{bd}
\end{eqnarray}
from $R_{\tau}=\Gamma(\tau\rightarrow e\nu\bar{\nu})
/\Gamma(\tau\rightarrow \mu\nu\bar{\nu})$ or 
$R_{\tau\mu}=\Gamma(\tau\rightarrow \mu\nu\bar{\nu})
/\Gamma(\mu\rightarrow e\nu\bar{\nu})$\cite{bound}, 
$|\lambda_{323}|<0.070\times(\tilde{m}/100GeV)$ becomes negligible 
while $|\lambda_{322}|$ can be as large as order one 
for $\tilde{M}\equiv m_{\tilde{\mu}}=20$~TeV, which is a natural 
assumption in the effective SUSY. 

Consequently, it is experimentally viable to take the $\lambda_{322}$ 
dominant case among $R$ parity violating couplings such that 
\begin{eqnarray}
\Delta a^{R\hspace{-1.75mm}/}_\mu \approx
34.9\times 10^{-10}\bigg(\frac{100GeV}{\tilde{m}}\bigg)^2|\lambda_{322}|^2.
\end{eqnarray}
In Fig. 4, the above equation is plotted in the $|\lambda_{322}|$ 
and $\tilde{m}$ plane. 
Here we show the $1\sigma$(dark grey) and $2\sigma$(light grey) bands 
beyond the standard model prediction\cite{davier}. 
It is shown that most of the parameter space is consistent with the 
experimental 
bound (\ref{bd}) for the decoupling assumption $\tilde{M}=20$~TeV. 
Note that there exists 
a enough room for perturbative $\lambda_{322}$ to explain the
BNL experiment even for a relatively large slepton mass 
in the third generation, 
for example
$\tilde{m}\sim 500$ GeV can give a perturbative
$\lambda_{322}$ coupling within the shaded region. 
If we consider the other terms in Eqs.~(\ref{l}) and 
(\ref{lp}) in the phenomenologically allowed region with
positive contributions, the $R$ parity 
violating couplings can be even lowered. [In this regard,
we note that the SUSY calculation of $a_\mu$ with
$\tan\beta\sim 40$ \cite{susy} amounts to the Yukawa couplings 
of order 1.]

In conclusion, we have shown that the effective 
supersymmetric model toward
a flavor conserving phenomena for the first two light
generations and the introduction of $R$ parity
violating couplings open up the possibility of a large
anomalous magnetic moment of $\mu$ within the accessible
range of the BNL experiment~\cite{bnl} and a discovery
potential of the third generation sfermions in the 
TevatronII and LHC accelerators.

\acknowledgments
We thank E. J. Chun and P. Ko for useful discussions.
This work is supported in part by the BK21 program of Ministry 
of Education, Korea Research Foundation Grant No. KRF-2000-015-DP0072, 
and by the Center for High Energy Physics(CHEP),
Kyungpook National University, and the Office of Research
Affairs, Seoul National University.

\end{document}